\documentclass[10pt]{article}

\usepackage{amsfonts}
\usepackage{amsmath}
\usepackage{amssymb}
\usepackage{compozition}
\usepackage{bm}
\usepackage{graphicx}
\usepackage[all]{xy}
\usepackage{MnSymbol}
\usepackage{mathrsfs} 
\usepackage{comment}
\usepackage{stackengine}

\newcommand{\squet}[1]{\left\vert #1 \right]}
\newcommand{\squar}[1]{\left[ #1 \right\vert}

\newcommand{\qtensr}[1]{\overset{\pmb{\   \neg}}{\mathbf{#1}}}
\newcommand{\qtensl}[1]{\overset{\pmb{\backneg}}{\mathbf{#1}}}
\newcommand{\squop}[1]{\overset{\pmb{\backneg \hspace{-2pt} \neg}}{\mathbf{#1}}}

\newcommand{\shortlongvert}{\hspace{0.3ex}\rule[-0.2ex]{0.3pt}{1.7ex}\hspace{0.21ex} \rule[-0.6ex]{0.3pt}{2.5ex}\hspace{0.3ex}}
\newcommand{\shortlongverttextmode}{\hspace{0.1ex}\rule[-0.2ex]{0.3pt}{1.7ex}\hspace{0.21ex} \rule[-0.6ex]{0.3pt}{2.5ex}\hspace{0.1ex}}

\newcommand{\SubseteqX}{\raisebox{1.3pt}{\stackunder[1.2pt]{$\Subset$}{\hspace{1pt}\rule[1pt]{.829ex}{.06ex}}}}

\tikzstyle{bigboxP} =[draw, fill=yellow!30,text depth = 4.5cm,minimum width=4cm,font=\Large]
\tikzstyle{bigboxE} =[draw, fill=yellow!30,text depth = 3cm,minimum width=4cm,font=\Large]
\tikzstyle{bigboxM} =[draw, fill=yellow!30,text depth = 4.5cm,minimum width=4cm,font=\Large]
\tikzstyle{bigboxT} =[draw, fill=yellow!30,text depth = 3cm, minimum width=4cm, text badly ragged, font=\Large]
\tikzstyle{smallbox}=[draw,fill=white, text width=2.5cm, text badly ragged]
\tikzstyle{bigboxQP} =[draw, fill=yellow!30,text depth = 5.5cm,minimum width=4cm,font=\Large]
\tikzstyle{bigboxQM} =[draw, fill=yellow!30,text depth = 5.5cm,minimum width=4cm,font=\Large]

\newcommand{\FlowchartlineWW}[4][0,0]{
\path[draw, rounded corners=12pt, ->]
let \p1=(#1), \p2=(#2.west), \p3=(#3.west)  in
(\x2, \y2+\x1) -- (\x2-#4cm, \y2+\x1) -- (\x2-#4cm,\y3+\y1) -- (\x3, \y3+\y1);
}

\newcommand{\FlowchartlineEE}[4][0,0]{
\path[draw, rounded corners=9pt, ->]
let \p1=(#1), \p2=(#2.east), \p3=(#3.east)  in
(\x2, \y2+\x1) -- (\x2+#4cm, \y2+\x1) -- (\x2+#4cm,\y3+\y1) -- (\x3, \y3+\y1);
}

\newcommand{\FlowchartlineWE}[4][0,0]{
\path[draw, rounded corners=9pt, ->]
let \p1=(#1), \p2=(#2.west), \p3=(#3.east)  in
(\x2, \y2+\x1) -- (\x2-#4cm, \y2+\x1) -- (\x2-#4cm,\y3+\y1) -- (\x3, \y3+\y1);
}

\newcommand{\FlowchartlineEW}[4][0,0]{
\path[draw, rounded corners=9pt, ->]
let \p1=(#1), \p2=(#2.east), \p3=(#3.west)  in
(\x2, \y2+\x1) -- (\x2+#4cm, \y2+\x1) -- (\x2+#4cm,\y3+\y1) -- (\x3, \y3+\y1);
}

\newcommand{\FlowchartlineEWred}[4][0,0]{
\path[draw, rounded corners=9pt, ->,thick, red]
let \p1=(#1), \p2=(#2.east), \p3=(#3.west)  in
(\x2, \y2+\x1) -- (\x2+#4cm, \y2+\x1) -- (\x2+#4cm,\y3+\y1) -- (\x3, \y3+\y1);
}

\newcommand{\FlowchartlineWEsmooth}[4][0,0]{
\path[draw, ->]
let \p1=(#1), \p2=(#2.west), \p3=(#3.east)  in
(\x2, \y2+\x1)  to [out=180, in=0]  (\x3, \y3+\y1);
}

\newcommand{\FlowchartlineEWsmooth}[4][0,0]{
\path[draw, ->]
let \p1=(#1), \p2=(#2.east), \p3=(#3.west)  in
(\x2, \y2+\x1)  to [out=0, in=180]  (\x3, \y3+\y1);
}

\newcommand{\FlowchartlineEWsmoothred}[4][0,0]{
\path[draw, ->, red, thick]
let \p1=(#1), \p2=(#2.east), \p3=(#3.west)  in
(\x2, \y2+\x1)  to [out=0, in=180]  (\x3, \y3+\y1);
}

\title{\textbf{Implementation of the Quantum Equivalence Principle}} 
\author{Lucien Hardy\\
\textit{Perimeter Institute,}\\
\textit{31 Caroline Street North,}\\
\textit{Waterloo, Ontario N2L 2Y5, Canada}}
\date{}

\begin{document}

\maketitle

\begin{abstract}
The quantum equivalence principle \cite{hardy2018construction} says that, for any given point, it is possible to find a quantum coordinate system with respect to which we have definite causal structure in the vicinity of that point.   It is conjectured that this principle will play a similar role in the construction of a theory of Quantum Gravity to the role played by the equivalence principle in the construction of the theory of General Relativity.

To actually implement the quantum equivalence principle we need a suitable notion of quantum coordinate systems - setting up a framework for these is the main purpose of the present paper.  First we introduce a notion of \emph{extended states} consisting of a superposition of terms (labeled by $u$) where each term corresponds to a manifold, $\mathcal{M}_u$,  with fields defined on it.  A quantum coordinate system consists of an identification of points between some subsets, $\mathcal{O}_u\subseteq \mathcal{M}_u$, of these manifolds along with a coordinate, $x$, that takes the same value on those points identified.

We also introduce a notion of quantum coordinate transformations (which can break the identification map between the manifolds) and show how these can be used to attain definite causal structure in the vicinity of a point.

We discuss in some detail how the quantum equivalence principle might form a starting point for an approach to constructing a theory of Quantum Gravity that is analogous to way the equivalence principle is used to construct General Relativity.
\end{abstract}

\section{Introduction}\label{sec:introduction}

The equivalence principle of Einstein can be stated in the following way
\begin{quote}
\textbf{The Equivalence Principle:} For any given point it is possible to find a coordinate system with respect to which we have inertial behavior in the vicinity of that point.
\end{quote}
Note that the fact that we have inertial behaviour means that the coordinate system represents free fall motion locally.  The power of the equivalence principle stems from the fact that it forms a bridge between pre-general relativistic physics (in which we can work in a global inertial reference frame) and General Relativity (in which we do not have a global inertial reference frame).   In Sec.\ \ref{sec:generalrelativity} we look in more detail at the conceptual structure within which General Relativity is obtained and the equivalence principle's role in this  (see Fig.\ \ref{fig:GR} in particular).

In a theory of Quantum Gravity we expect to have indefinite causal structure (as we will have something like a quantum superposition of different solutions for the metric). If we take definite causal structure to be analogous to inertial behaviour then this suggests the following principle which, I hope, will play a similar role to the equivalence principle but in guiding us to a theory of Quantum Gravity.
\begin{quote}
\textbf{The Quantum Equivalence Principle:} For any given point it is possible to find a quantum coordinate system with respect to which we have definite causal structure in the vicinity of that point.
\end{quote}
To implement the quantum equivalence principle we will need to find an appropriate notion of \emph{quantum coordinate systems}.  In Sec.\ \ref{sec:quantumgravity} we consider a possible conceptual structure for obtaining Quantum Gravity (see Fig.\ \ref{fig:QG} in particular) that is analogous to that of General Relativity and look at how the quantum equivalence may play an analogous role to that of the equivalence principle in General Relativity.

The quantum equivalence principle, which I proposed in \cite{hardy2018construction}, was strongly motivated in the first place by the work of Giacomini, Castro-Ruiz, and Brukner on \emph{quantum reference frames} \cite{giacomini2019quantum}.  This provides a notion of quantum reference frames at a given time that can correspond to a superposition of other frames of reference (at that given time).  This principle is supported by the work by Guerin and Brukner on causal reference frames \cite{guerin2018observer}, and related work by Oreshkov \cite{oreshkov2018whereabouts}.  In these works it is shown how physically equivalent circuit representations can have event $A$ localized in time while event $B$ is delocalized or vice versa.  This can be regarded as a discrete example of the quantum equivalence principle.

In the present work we seek a notion of quantum coordinates systems that is analogous to the notion of coordinate systems used in General Relativity.  In General Relativity a coordinate system labels the points in some region of the manifold.  This labeling is conceptually prior to the introduction of the metric and so the coordinate system knows nothing of space, time, and causal structure as such.  Hence, we need to go beyond the ideas of Giacomini {\it et al.} which operate at a given time.  We also expect these coordinates to be continuous.  Hence we also need to go beyond the discrete circuit framework of Guerin and Brukner and of Oreshkov.

The notion of quantum coordinate systems proposed here is both continuous and operates at the space-time level.

Many authors have proposed quantum equivalence principles of various sorts \cite{castagnino1980quantum, candelas1983there, osipova1986quantum, hessling1994quantum, kleinert1997quantum, jejjala2007there, caticha2011entropic} that are quite different from the one given above (generally these are still concerned with inertial motion of some sort).   Interesting work on the conceptual role an appropriate (though undefined) quantum equivalence principle might play as a bridge from Quantum Field Theory to Quantum Gravity has been undertaken by Pipa {\it et al. }  There has also been research on applying Einstein's equivalence principle to Quantum Theory (this is a vast subject but see \cite{viola1997testing, altschul2015quantum, zych2018quantum, rosi2017quantum} for some work along this direction).

\section{Basic idea in this paper}

The basic idea to implement the quantum equivalence principle is as follows.  First we define $u$ as
\[ u= \left\{ (\Phi, p) : \forall p\in\mathcal{M}_u \right\}  \]
where $\Phi$ are the set of tensor fields that define the physics (corresponding to matter fields and the metric field).  Thus, $u$, is a specification of a classical state. We write $\squet{u}$ to describe the situation $u$ (we could have written $|u\rangle$ but we need our $\squet{u}$ notation to do some extra things that make it worthy of new notation).   Next we argue from the path integral approach that the object
\[ \squet{\Psi}  = \int \mathcal{D}u~  c_u \squet{u}    \]
(which we call an \emph{extended state})  contains the necessary physical information (actually we consider more general objects called \emph{extended $\mathbf{A}$-states} but for this overview it is sufficient to talk about extended states).

A quantum coordinate system is given by providing an identification between points in different $\mathcal{O}_u\subseteq \mathcal{M}_u$ for different $u$ and then providing a coordinate, $x$, for these points.  This identification is pure gauge and has no physical significance in itself.  We can imagine covering all points in $\{ \mathcal{M}_u\}$ by a set of charts employing quantum coordinate systems in this fashion (forming a quantum atlas).  A quantum diffeomorphism can change the identification mapping between the $\mathcal{O}_u$.   By applying a quantum diffeomorphism that keeps some point, $x$, fixed we can \lq\lq rotate" the solutions for the different $u$ so that the lightcones associated with the metrics align (we can also get the conformal factors to match).  In so doing the causal structure at $x$ becomes definite.

The idea of identifying points across different manifolds in a quantum superposition has been independently arrived at by Ding Jia \cite{jia2019quantum, jia2019causally} with different motivations in mind.   In particular, points are identified that share some common features (such as fields taking a given value).    Here we do not impose this constraint but rather regard the identification as pure gauge (and different identifications are equally valid).   Applications Jia considers applications for his identification structure are to ensure that all terms in the superposition have the same boundary conditions in a Feynman path integral and also to help in collecting terms together in a useful way when evaluating these path integrals.

\section{Background}

In \cite{hardy2005probability, hardy2007towards, hardy2009quantum3} I set up a general operational framework (the causaloid framework) for the purpose of studying indefinite causal structure.   This was in the spirit of the general probability theories framework \cite{hardy2001quantum, barrett2007information}.   Since then two teams, Chiribella, D'Ariano, Perinotti, and Valiron (CDPV) \cite{chiribella2009beyond} working initially in Pavia, and Oreshkov, Costa, and Brukner (OCB)\cite{oreshkov2012quantum} working initially in Vienna have developed operational frameworks based on quantum operators for studying indefinite causal structure.  All three approaches (the causaloid and the operator based approaches) have in common that they are linear in probability and that they associate state-like objects with what are effectively arbitrary shaped regions of space-time.  The operator tensor framework \cite{hardy2011reformulating, hardy2012operator}, which is a development of the causaloid approach to Quantum Theory, can also be applied to indefinite causal structure (see \cite{hardy2018foundations} for a treatment of OCB's approach).   CDPV proposed the idea of a quantum switch - an explicit way to obtain indefinite causal structure.  And OCB proposed some inequalities which, if violated, demonstrate indefinite causal structure. Recent work (already mentioned in Sec.\ \ref{sec:introduction})  by Oreshkov \cite{oreshkov2018whereabouts} and also Guerin and Brukner \cite{guerin2018observer} introduces a new perspective on indefinite causal structure.  The field of indefinite causal structure has grown considerably in recent years and there are many more papers beyond those cited above.

Robert Oeckl's general boundary formalism (both the original amplitude based version \cite{oeckl2003general} and the more recent postive formlism \cite{oeckl2013positive}) also connects with the work on indefinite causal structure.

Some of the ideas in this paper come from my operational formulation of General Relativity \cite{hardy2016operational}.  This is a reformulation of General Relativity as an operational probabilistic theory motivated by the problem of Quantum Gravity (which will likely be both operational and probabilistic).  In particular, the representation of classical configurations as $u= \left\{ (\Phi, p) : \forall p\in\mathcal{M}_u \right\}$ was introduced in that paper (though they were denoted by $\Psi$ rather than $u$).  The notion of \lq\lq chartable space" as a place where different manifolds can live was introduced there and would find application to the ideas presented here.

The field of quantum reference frames goes back to the 1967 with a paper by Aharonov and Susskind \cite{aharonov1967charge} on charge superselection, followed in 1984 with a paper by Aharonov and Kaufherr \cite{aharonov1984quantum} entitled \lq\lq Quantum frames of reference".   A second wave of activity happenend a dozen years ago (see the review paper \cite{bartlett2007reference}) using more modern notions from quantum information.   Most recently the above mentioned paper \cite{giacomini2019quantum} by Giacomini {\it et al}.\  has reignited the field.  Vanrietvelde, H\"ohn, Giacomini, and Castro-Ruiz \cite{vanrietvelde2018change, vanrietvelde2018switching} have explored how to switch between reference frames via a \lq\lq perspective-neutral" framework.  H\"ohn has explored how quantum reference frames may impact on the study of Quantum Gravity \cite{hoehn2018quantum}.

The present paper is an elaboration on ideas I first presented in \cite{hardy2018construction} where the quantum equivalence principle was first stated and, further, the idea that we might pursue a route to Quantum Gravity analogous to Einstein's development of General Relativity was considered.

We will make much use of the path integral due originally to Feynman.  The path integral has frequently been invoked for looking at the problem of Quantum Gravity (see \cite{bunster2017path} for recent perspectives) forming the basic motivation for the causal sets approach  \cite{bombelli1987space}, the spin foam approach \cite{perez2013spin}, the dynamical triangulations approach \cite{ambjorn1997geometry}, the Euclidean gravity approach \cite{gibbons1993euclidean} as well as various cosmological models \cite{feldbrugge2017lorentzian}.

\section{Path from the path integral to extended states for a particle}\label{sec:PItoESparticle}

To get started consider the simple case of the path integral for a single particle moving in one dimension.  In this case the amplitude for starting at $y_i$ at time $t_i$ and arriving at $y_f$ at time $t_f$ is given  by Feynman's path integral
\begin{equation}\label{pathintegral}
\frac{1}{Z}\int_{u\in V[y_i,t_i,y_f, t_f]}\mathcal{D}u~ e^{iS/\hbar}
\end{equation}
where
\begin{equation}
u= \{ \left(y(t),t\right): t_i \leq t \leq t_f \}
\end{equation}
This is a path starting at time $t_i$ and ending at time $t_f$.   We will usually denote such paths by $u$ but we will also use $v$ and $w$ for these paths when we need extra symbols.   $\mathcal{D}u$ is a measure over such paths so we can perform a functional integral.   $V[y_i,t_i,y_f, t_f]$ is the set of paths that start at $(y_i,t_i)$ and end at $(y_f,t_f)$. $Z$ is a normalization factor. $S$ is the action given by integrating the Lagrangian along a path $u\in V[y_i,t_i,y_f, t_f]$
\begin{equation}
S[y_i,t_i,y_f, t_f] = \int_{u\in V[y_i,t_i,y_f, t_f]} dt L(t)
\end{equation}
In the expression \eqref{pathintegral} we have integrated over paths and lost information accrued along each individual path. We will consider a series of mathematical objects which contain successively more information until we obtain an object that is useful for the purposes we have in mind.

As a first step, consider the object,
\begin{equation}\label{stepone}
\squet{\Psi_S^{V[t_i, t_f]}}= \int_{u\in V[t_i,t_f]} \mathcal{D}u~ \frac{1}{Z} e^{iS/\hbar} \squet{u}
\end{equation}
where $\squet{u}$ is a way to refer to the path $u$ in a manner analogous to the way the notation $|x\rangle$ refers to the position $x$ (we will elaborate on this notation later). Now we sum over all paths starting (at any position) at time $t_i$ and ending at time $t_f$.  We can construct the path integral in \eqref{pathintegral} from the object in \eqref{stepone} as follows
\begin{equation}\label{psiStoPI}
\int_{u\in V[y_i,t_i,y_f,t_p]} \mathcal{D}u~  e^{iS/\hbar} = \int_{v\in V[y_i,t_i,y_f,t_p]} \mathcal{D}v~ \left[v\right. \squet{\Psi_S^{(t_i, t_f)}}   
\end{equation}
where we put
\begin{equation}
\left[ v|u \right] = \delta(v-u)  ~~~~ \text{where}~~~~ \int_u \mathcal{D}u~ \delta(v-u) f(u) = f(v)
\end{equation}
The object in \eqref{stepone} keeps the information accrued along each individual path.  Further, it is a sum over all paths for the given time interval, not just those starting at $y_i$ and ending at $y_f$.  However, it still contains a sum - the integral over the Lagrangian to obtain the phase.  This means (i) that we have lost information about the contribution to the phase coming from each infinitesimal section of the path (i.e.\ about the Lagrangian at each point) and (ii) we have committed to a particular time interval for each term.

Hence, we consider the object
\begin{equation}
\squet{\Psi_L^V} = \int_{u\in V} \mathcal{D}u~\frac{1}{Z} L_u(\mathcal{N}_u) \squet{u}
\end{equation}
We call this an \emph{extended} $L$-\emph{state}.  Here $\mathcal{N}$ is a time interval (more generally, it may be the union of a number of non-overlapping time intervals),
\begin{equation}
u = \left\{ (y,t): t\in \mathcal{N}_u \right\}
\end{equation}
and
\begin{equation}
L_u(\mathcal{N}_u) = \left\{ (L,t): t\in \mathcal{N}_u \right\}
\end{equation}
Importantly note that we allow the coefficient in front of the $\squet{u}$ to be a set comprised of elements $(L,t)$ for a times $t\in\mathcal{N}$.  This allows us to maintain information about the contribution to the phase coming from each part of the path.  We also generalize and allow paths for arbitrary $\mathcal{N}$.  The set, $V$, of paths we consider may be arrived at by various means in this more general setting. For the moment we simply note that we do not require that every entry  have the same $\mathcal{N}$.

Next we will look at how the path integral can be obtained from $\squet{\Psi_L^V}$.  First we must provide a more general notion than $[u|v]$ applying more generally to situations when $u$ and $v$ may pertain to different $\mathcal{N}$.  We need this because the path $u$ may be contained in the path $v$ (where $\mathcal{N}_u \subset \mathcal{N}_v$):
\begin{equation}
\text{If} ~~u\subseteq v ~~~ \text{then}~u ~\text{contained in}~v
\end{equation}
To clarify, $u\subseteq v$ means that
\begin{equation}
u=v_{\mathcal{N}_u} ~~~\text{where} ~~~v\vert_{\mathcal{N}_u} = \left\{ (y, t)\in v: \forall t\in\mathcal{N}_u \right\}
\end{equation}
Now we define
\begin{equation}
[ u \shortlongvert v ] = \delta(u -  v\vert_{\mathcal{N}_u})
\end{equation}
Note that this is asymmetric (and we have indicated this using $\shortlongverttextmode$ which is supposed to be  indicative of the example where a longer path gets shortened).  If $u$  is not contained in $v$ then $[u\shortlongverttextmode v]=0$.    We also define
\begin{equation}
[ u \shortlongvert A_v(\mathcal{N}_v)| v ] = A_u(\mathcal{N}_u) \delta(u -  v\vert_{\mathcal{N}_u})
\end{equation}
This is non-zero only when $u$ is contained in $v$, and in this case we keep only the elements of $A_v(\mathcal{N}_v)$ pertaining to elements in $\mathcal{N}_u$.

We will say
\begin{equation}
W \SubseteqX V ~~\iff ~ \forall~  u\in W ~ \exists ~ v\in V ~\text{such that} ~ u\subseteq v
\end{equation}
If $W\SubseteqX V$ then each path in $W$ is contained in some path in $V$.

In the case where $V[t_i, t_f]\SubseteqX V$, we can use $\squet{\Psi_L^V}$ to calculate $\squet{\Psi_{S[y_i,t_i,y_f, t_f]}}$ as follows. First we project
\begin{equation}\label{projecttoPsiL}
\squet{\Psi_L^{V[t_i, t_f]}} = \int_{v\in V[t_i, t_f]}\mathcal{D}v \squet{v} [v\shortlongvert ( \squet{\Psi_L^V})
\end{equation}
Then we integrate and exponentiate
\begin{equation}
\squet{\Psi_S^{V[t_i, t_f]}} =  (\exp\text{int})_{op} \squet{\Psi_L^{V[t_i, t_f]}}
\end{equation}
We use the notation $(\exp\text{int})_{op}$ to indicate that we have an operator acting linearly on each term in $\squet{\Psi_L^{V[t_i, t_f]}}$ that integrates over elements in the set, $L_u(\mathcal{N}_u)$, and then exponentiates.  Once we have $\squet{\Psi_S^{(t_i, t_f)}}$ we can use \eqref{psiStoPI} to get the amplitude.  Note that we could simply have projected down to $\squet{\Psi_L^{V[x_i, t_i, x_f, t_f]}}$ directly in \eqref{projecttoPsiL} and then we only require that $V[x_i, t_i, x_f, t_f]\SubseteqX V$.  We choose the slightly more complicated route here for pedagogical reasons as we will set up a similar approach for Quantum Gravity in the next section.

Finally, we are not restricted to having the Lagrangian in the coefficient. More generally, we can consider extended $A$-states
\begin{equation}
\squet{\Psi_A} = \int \mathcal{D}u~ c_u A_u(\mathcal{N}_u) \squet{u}
\end{equation}
where
\begin{equation}
A_u(\mathcal{N}_u) = \left\{ (A,t): t\in \mathcal{N}_u \right\}
\end{equation}
for the quantity $A$ calculable at each time $t$.  Also we can, in general, let the coefficient $c_u$ depend on $u$.   One reason for allowing such a dependence is that some paths may be blocked or be partially absorbed.  We have dropped the superscript, $V$, in $\squet{\Psi_A}$ as now we can regard this as integration over all $u$ with $c_u$ equal to zero for $u$ not in the given set.

It is worth noting that the object
\begin{equation}\label{extendedstate}
\squet{\Psi} = \int \mathcal{D}u~c_u \squet{u}
\end{equation}
actually contains all the information we need.  We can set up a linear operator that acts on each $\squet{u}$ and brings out $A_u(\mathcal{N}_u)$ as a coefficient in front of this term.  We will call the object in \eqref{extendedstate} the \emph{extended state}.

\section{From the path integral to extended states in Quantum Gravity}

We can set up a similar path integral for Quantum Gravity.  First we define
\begin{equation}\label{QGu}
u = \left\{ (\Phi, p): \forall p\in\mathcal{M}_u \right\}
\end{equation}
where $\Phi$ is a list of all the fields - the matter fields plus the metric field (so $\Phi$ is playing an analogous role to $y$ of Sec.\ \ref{sec:PItoESparticle}).  Further, $\mathcal{M}_u$ is a subset of manifold points (analogous to $\mathcal{N}_u$ of Sec.\ \ref{sec:PItoESparticle} which was the union of disjoint time intervals). Typically we would like $\mathcal{M}_u$ to be well behaved. We will assume it is union of disjoint \emph{manifolds with corners} \cite{melrose1996differential, joyce2009manifolds}.  Roughly speaking, a manifold with corners is one that can everywhere be locally covered by the points in $[0,\infty)^k \mathbb{R}^{N-k}$ - this allows it to have boundaries and corners (in the vicinity of these $k$ is non-zero).  See \cite{hardy2016operational} for more motivation for using such manifolds with corners.  When we refer to a \lq\lq manifold" in what follows we are referring to these well behaved subsets of manifolds (the union of disjoint manifolds with corners) which are, in any case, a generalisation of the usual notion of manifolds.

Consider a boundary, $b$, at which we can impose boundary conditions $a$.  We need not go into detail for the time being except to note that physically meaningful boundaries and boundary conditions should be invariant under diffeomorphisms (and, more generally, under the quantum version of such diffeomorphisms).  A treatment of boundary conditions invariant under diffeomorphisms appropriate to this situation is given in \cite{hardy2016operational}.  In Sec.\ \ref{sec:quantumdiffeomorphisms} we treat quantum diffeomorphisms and in Sec.\ \ref{sec:beables} we say what the beables are.  Boundary conditions must be beables in this sense.   We can now calculate the amplitude associated with the given boundary conditions as
\begin{equation}\label{QGPI}
  \int_{u\in V[a]} \mathcal{D}_u e^{iS/h}
\end{equation}
Here $V[a]$ is the set of all $u$ consistent with boundary conditions, $a$, are analogous to $(y_i, t_i, y_f, t_f)$ from Sec.\ \ref{sec:PItoESparticle} and the boundary, $b$, is analogous to $(t_i, t_f)$.

As before, we can consider the more general object
\begin{equation}
\squet{\Psi_S^{V[b]}} = \int_{u\in V[b]} \mathcal{D}u ~ \frac{1}{Z} e^{iS/\hbar} \squet{u}
\end{equation}
where $V[b]$ is the set of all $u$ consistent with any boundary condition at $b$.  We can use $\squet{\Psi_S^{V[b]}}$ to  calculate the amplitude in \eqref{QGPI} using
\begin{equation}\label{QGpsiStoPI}
\int_{u\in V[a]} \mathcal{D}u~  e^{iS/\hbar} = \int_{v\in V[a]} \mathcal{D}v~ \left[v\right. \squet{\Psi_{S[\{a\}]}}
\end{equation}
This object contains integrals (in the form of $S$) and also commits to a certain boundary condition.

We can consider a more general object still - this is the extended $L$-state for Quantum Gravity:
\begin{equation}\label{PsiLVQG}
\squet{\Psi_L^V} = \int_{u\in V} \mathcal{D}u~\frac{1}{Z} L_u(\mathcal{M}_u) \squet{u}
\end{equation}
where
\begin{equation}
L_u(\mathcal{M}_u) = \left\{ (L, p): \forall p\in \mathcal{M}_u \right\}
\end{equation}
Note that we allow $V$ in \eqref{PsiLVQG} to be more general than $V[b]$.  The manifolds, $\mathcal{M}_u$, pertaining to $u\in V$ do not have to pertain to the region defined by some boundary conditions $b$.  They can be any (reasonably well behaved) manifolds.

As in Sec.\ \ref{sec:PItoESparticle}, we can set up notions that allow us to project $\squet{\Psi_L^V}$.  First we say that $u$ is contained in $v$ iff $u\subseteq v$.  This is equivalent to saying that $u=v\vert_{\mathcal{M}_u}$ where
\begin{equation}
v\vert_{\mathcal{M}_u} = \left\{ (\Phi, t)\in v: \forall t\in\mathcal{M}_u \right\}
\end{equation}
We define
\begin{equation}
[ u \shortlongvert v ] = \delta(u -  v\vert_{\mathcal{M}_u})
\end{equation}
and
\begin{equation}
[ u \shortlongvert A_v(\mathcal{M}_v)| v ] = A_u(\mathcal{M}_u) \delta(u -  v\vert_{\mathcal{M}_u})
\end{equation}
Finally, we say $W\SubseteqX V$ iff, for each $u\in W$, there exists a path $v\in V$ that contains $u$.

If $V(b)\SubseteqX V$ then we can project (just as we did in \eqref{projecttoPsiL}) obtaining
\begin{equation}
\squet{\Psi_L^{V[b]}} = \int_{v\in V[b]}\mathcal{D}u \squet{v} [v\shortlongvert ( \squet{\Psi_L^V})
\end{equation}
Now we can obtain $\squet{\Psi_S^{V[b]}}$ by applying the integration and exponentiation operator
\begin{equation}
\squet{\Psi_S^{V[b]}} =  (\exp\text{int})_{op} \squet{\Psi_L^{V[b]}}
\end{equation}
From this we can use \eqref{QGpsiStoPI} to calculate the amplitude associated with a particular boundary condition $a$ via the path integral.

As before we can consider other objects. A more general object than $\squet{\Psi_L^V}$ is the extended $\mathbf{A}$-state:
\begin{equation}
\squet{\Psi_\mathbf{A}} = \int \mathcal{D}u~ c_u \mathbf{A}_u(\mathcal{M}_u) \squet{u}
\end{equation}
where $\mathbf{A}$ can be any tensor field (the Lagrangian is then a special case).  We have dropped the $V$ superscript as this is superseded by the $c_u$ dependence.  Also, we can start with the \emph{extended state}
\begin{equation}
\squet{\Psi} = \int \mathcal{D}u ~ c_u  \squet{u}
\end{equation}
Acting on this with linear operators can be made to bring out the information we need to calculate the other objects.

\section{Quantum diffeomorphisms}\label{sec:quantumdiffeomorphisms}

A classical diffeomorphism, $\varphi$, is a smooth invertible map on the manifold taking $p$ to $\varphi(p)$.   This induces transformations on tensor fields that live on the manifold which we denote by saying that $\mathbf{A}(p)$ goes to $\varphi^*\mathbf{A}(\varphi(p))$.    If we act on $u$ with a diffeomorphism then we have
\begin{eqnarray*}
  \varphi^*u &=& \left\{ (\varphi^*\mathbf{A}, \varphi(p)) : \forall \varphi(p) \in \varphi(\mathcal{M}) \right\} \\
                    &=& \left\{ (\varphi^*\mathbf{A}, p) : \forall p \in \varphi(\mathcal{M}) \right\}
\end{eqnarray*}
where the second line follows as $\varphi(p)$ is a dummy variable in the first line.

We will first define a restricted notion of a quantum diffeomorphism.  We will use this to motivate the general definition.
The basic action of a restricted quantum diffeomorphism is to perform a classical diffeomorphism on each of the different $\squet{u}$ terms.   A restricted quantum diffeomorphism is a linear operator, $\squop{\varphi}$, defined by a set of classical diffeomorphisms, $\{ \varphi_u: \forall u\}$  that acts on $\squet{\Psi_\mathbf{A}}$ as follows
\begin{eqnarray}
\squop{\varphi} \squet{\Psi_\mathbf{A}}  &=& \squop{\varphi}  \int  \mathcal{D}u~ c_u \mathbf{A}_u(\mathcal{M}_u) \squet{u} \\
   &=& \int  \mathcal{D}u~ c_u e^{i\theta(u,{\varphi_u}^*u)}{\varphi_u}^*\mathbf{A}(\varphi_u(\mathcal{M}_u)) \squet{{\varphi_u}^* u}
\end{eqnarray}
where
\begin{equation}
{\varphi_u}^*\mathbf{A}(\varphi(\mathcal{M}_u)) = \left\{ ({\varphi_u}^*\mathbf{A}, p): \forall p\in \varphi(\mathcal{M}) \right\}
\end{equation}
and
\begin{equation}
\theta(u, v) = - \theta(v,u)
\end{equation}
This phase function is some given function prescribed by the theory (its form may, though, depend on what type of tensor field $\mathbf{A}$ is). It need only be defined for arguments $u$ and $v$ which can be transformed into one another by a diffeomorphism.  We require it to be anti-symmetric so that if we find a quantum diffeomorphism that undoes the action of a previous quantum diffeomorphism, then $\squet{\Psi_A}$ returns to its original version without accumulating any phase factors.  One possible choice is
\begin{equation}
\theta(u,v)=0
\end{equation}
as this is antisymmetric.  At least in the case where $\mathbf{A}$  is a scalar field we are motivated to choose $\theta(u,v)=0$ since then we want the extended $L$-state, $\squet{\Psi_L}$, to have real coefficients before and after a change of quantum coordinates so that we get the correct path integral.  In general, however, there may be physical reasons for some other choice.  In particular, there is a non-trivial phase associated with the quantum frame of reference change in the work of Giacomini {\it et al.}.  It may be a challenge to find physically meaningful functions of $u$ and $v$.
These restricted quantum diffeomorphisms are not necessarily invertible.  This is because we could have ${\varphi_u}^* u ={\varphi_{v}}^* v$ (where $u$ and $v$ are different).  This does not matter as, by virtue of the way we will set the theory up,  no physical information is lost under quantum diffeomorphisms.

We will now give the general definition of a quantum diffeomorphism.  These also are not invertible.  However, as we will see, the action of a quantum diffeomorphism on any given $\squet{\Psi_\mathbf{A}}$ can be reversed by an appropriate choice of quantum diffeomorphism.   A quantum diffeomorphism, $\squop{\varphi}$, is given by a set
\begin{equation}
\left\{ \sum_\alpha a_\alpha \varphi^\alpha_u: \forall u \right\}
\end{equation}
where $a_\alpha$ is real for all alpha and $\sum_\alpha a_\alpha^2 =1$.  The sum over the label $\alpha$ could instead be an integral.  The action of $\squop{\varphi}$ on $\squet{\Psi_\mathbf{A}}$ is as follows
\begin{equation}
\squop{\varphi} \squet{\Psi_\mathbf{A}}  = \int  \mathcal{D}u~ c_u  \sum_\alpha a_\alpha e^{i\theta(u,{\varphi^\alpha_u}^*u)} {\varphi^\alpha_u}^*\mathbf{A}(\varphi_u(\mathcal{M}_u)) \squet{{\varphi_u}^* u}
\end{equation}
It is clear now that, although quantum diffeomorphisms are not invertible, their action on a given $\squet{\Psi_\mathbf{A}}$ can be inverted.  Since quantum diffeomorphisms are not invertible, they do not form a group.  They do, however, form a monoid (that is a semigroup with an identity element).

Just as classical General Relativity is invariant under diffeomorphisms, Quantum Gravity is, we propose, invariant under quantum diffeomorphisms.   This has particular significance for the ontologically real quantities in the theory.


\section{Beables}\label{sec:beables}

The term \emph{beables} was coined by John Bell to refer to the ontologically real quantities in a theory.

In General Relativity, the beables are those functions of solutions that are invariant under diffeomorphisms (see discussion in \cite{hardy2016operational}).  Thus, if the world is described by $u$ then it is equally well described by $\varphi^* u$ for any $\varphi$ and beables are functions, $B(u)$, having the property that
\begin{equation}
B(u) = B(\varphi^* u) ~~~~~~\forall \varphi
\end{equation}
We cannot talk about the beables in a particular patch of the manifold as a diffeomorphism will replace the fields living there with different fields.  Thus, beables, in General Relativity are in some sense nonlocal.

What are the beables in Quantum Gravity?  Here we propose that they are quantities that are invariant under quantum diffeomorphisms. Thus, $B(\squet{\Psi})$ is a beable if and only if
\begin{equation}
B( \squet{\Psi}) = B(\squop{\varphi}\squet{\Psi}) ~~~~~~\forall \squop{\varphi}
\end{equation}
This means that $\squop{\varphi}\squet{\Psi}$ contains the same physical information as $\squet{\Psi}$.

A natural question that arises here is whether this way of formulating beables will offer any resolution to the interpretational problems of quantum theory.  In particular, will it resolve the measurement problem and will it offer some way of understanding Bell-type nonlocality?    A good way to develop an intuition with regard to these questions would be to look at some examples.  For the time being we can note that transformations between different quantum coordinate systems can remove superpositions and entanglement (this is true in the scheme of Giacomini {\it et al}.\ \cite{giacomini2019quantum} and will also hold here).  This offers some hope that macroscopic superpositions and entanglement based nonlocality are gauge artifacts and will not truly be present amongst the beables.

We can use beables to select on extended states.  For example, we may have seen a certain outcome, $\beta$, in some experiment, $E$, and the extended state must be consistent with this.  Then we have beable value
\begin{equation}
B_E( \squet{\Psi}) = \beta
\end{equation}
This constrains the extended state associated with this outcome.
One way to do this is using operational space as outlined in \cite{hardy2016operational}.  Operational space is given by nominating an ordered set of scalars, $\mathbf{S}=(S_1, S_2, S_3, \dots, S_K)$.  If we plot $u$ into operational space then we will obtain a surface that is $N$ dimensional at most (where $N$ is the dimension of spacetime).   We can select on $u$ that plot into certain regions, $A$, of operational space.  Then we require that the beables satisfy
\begin{equation}
B_\text{op space} (\squet{\Psi}) \subset A
\end{equation}
This means that we can only have terms, $\squet{u}$, in the expansion of $\squet{\Psi}$ that plot into $A$.    The discussion of operational space in \cite{hardy2016operational} is applicable here.

\section{Quantum coordinate systems}\label{sec:quantumcoordinates}

In General Relativity, coordinate systems are introduced to cover a manifold.  In our treatment of Quantum Gravity, the situation is complicated by the fact that we have a set of manifolds, $\{\mathcal{M}_u \}$, rather than a single manifold.   Nevertheless, we can set up a map that identifies points on different manifolds in $\{\mathcal{M}_u \}$ in some region.  We can then lay down coordinates.   We will call these \emph{quantum coordinate systems}.  This will be appropriate as we will be able to transform between them by means of a quantum transformation corresponding to a quantum diffeomorphism.  Indeed, a  diffeomorphism is the abstract version of a change of coordinate system.  Likewise, we will be able to think of a quantum diffeomorphism as the abstract version of a change of quantum coordinate system.

First, we choose a set of smooth invertible maps
\begin{equation}
\varphi_{u\rightarrow v}(p)
\end{equation}
for all $u$ and $v$  mapping
 \begin{equation}
 p\in\mathcal{O}_u\subseteq \mathcal{M}_u
 \end{equation}
 to
 \begin{equation}
 \varphi_{u\rightarrow v}(p)\in \mathcal{O}_{v} \subseteq \mathcal{M}_{v}
 \end{equation}
 such that
\begin{equation}
\varphi_{u\rightarrow w}(p) = \varphi_{v\rightarrow w} \circ \varphi_{u\rightarrow v}(p)   ~~~\forall ~ p\in \mathcal{O}_u
\end{equation}
so points are mapped consistently between $\{\mathcal{O}_u\}$.  We will call the maps $\varphi_{u\rightarrow v}$ \emph{identification maps}.   Note, the notation $\varphi_{u\rightarrow v}$ may be a little misleading. These maps do not map $u$ to $v$, but rather they map a subset of points (those in $\mathcal{O}_u$)  in the manifold associated with $u$ to a subset of points (those in $\mathcal{O}_v$) in the manifold associated with $v$.   We can always do this by choosing $\mathcal{O}_u$  that are coverable by points in $V_u\subseteq\mathbb{R}^N$. \footnote{A subtlety arises if some of the $\mathcal{O}_u$ include points in the boundary of $\mathcal{M}_u$.  Then the maps $\varphi_{u\rightarrow v}$ are only invertible for points in $\mathcal{O}_u$. In particular, it may then be that there are points in $\mathcal{O}_{v}$ that do not map to points in $\mathcal{O}_u$.  In this case we can simply map from some reference set, $\mathcal{O}$, to each of the $\mathcal{O}_v$.  For the most part, we will not concern ourselves with this subtlety.}

We can now cover these points, so identified, with coordinates, $x=(x^\mu: \mu=1, 2, \dots N)$.  To do this we can set up a bijection,  $x_u(p)$ from the points $p\in \mathcal{O}_{u}$ to the points in $V_{u}\subseteq \mathbb{R}^N$ for some particular $u$ then use the map $\varphi_{u\rightarrow v}$ to generate a coordinate system for every other element of $\{\mathcal{O}_u\}$.
\begin{equation}
x_{v}(p) = x_u(\varphi_{v \rightarrow u}(p))
\end{equation}
In this way we have laid down a coordinate system that covers part of each element of $\{\mathcal{M}_u\}$ identifying points with the same coordinates, $x$.

This quantum coordinate system only covers part of each manifold, $\mathcal{M}_u$.  This is analogous to the way a (classical) coordinate system (some times called a \emph{chart}) only covers part of a manifold.  In the classical case, a set of charts that cover the manifold is called an atlas.  We could consider setting up a \emph{quantum atlas} consisting of enough quantum coordinate systems (we could also call them \emph{quantum charts}) to cover every part of every $\mathcal{M}_u$.  In Sec.\ \ref{sec:quantummanifolds} we will discuss the tentative idea of \emph{quantum manifolds} which provides an implementation of a quantum atlas.

\section{Quantum coordinate transformations}\label{sec:quantumcoordinatetransformations}

We can perform a \emph{classical coordinate transformation} simply by transforming $x^\mu$ as in General Relativity.
\begin{equation}
x^\mu \rightarrow x^{\mu'} = f(\{ x^\mu\})
\end{equation}
This changes the name of the coordinate at each point. However, it does not break the identification map between the elements of $\{\mathcal{M}_u\}$.  A more radical transformation is a \emph{quantum coordinate transformation} which does break this identification map.   To do this we act on the maps $\varphi_{u\rightarrow v}$ with the elements of $\{\varphi_u\}$ (that can be used to define a restricted quantum diffeomorphism) to obtain a new identification map
\begin{equation}
\varphi_{u\rightarrow v} \rightarrow \varphi'_{u\rightarrow v}= \varphi_{v} \circ \varphi_{u\rightarrow v} \circ \varphi^{-1}_u
\end{equation}
between the elements in the set $\{ \mathcal{O}'_u\}$ (where  $\mathcal{O}'_u=\varphi_u(\mathcal{O}_u)$).  Each each element of this set is covered by coordinates $x'_u(p)= x_u(\varphi_u(p))$.  It is worth noting that the quantum aspect of the quantum coordinate system is in the identification between manifold points since this is something that can be changed by a quantum coordinate transformation (but not by a classical coordinate transformation).

A quantum coordinate system is simply attached to a set of manifolds (this is passive in that it does not change the extended state, $\squet{\Psi_A}$).  A transformation between different quantum coordinate systems is associated with a set of diffeomorphisms on each of these manifolds.  This set of diffeomorphisms can be used to implement a quantum diffeomorphism on an extended state, $\squet{\Psi_A}$.  What is the relationship between this quantum diffeomorphism with its action on an extended state and the quantum coordinate system?  To investigate this first note that we can take an active (rather than passive) approach. Thus, when we want to identify points in $\mathcal{O}_u$ and $\mathcal{O}_{v}$ we can perform an active transformation so that these points actually coincide.  We choose some particular $u=\tilde{u}$ (it does not matter which one) with an associated $\mathcal{O}_{\tilde{u}}$. Then we map all other $\mathcal{O}_u$ to overlap with $\mathcal{O}_{\tilde{u}}$ this by using the transformation maps $\varphi_{u\rightarrow \tilde{u}}$.   The transformation of $\squet{\Psi_A}$ is
\begin{equation}\label{PsiAoverlap}
\squet{\Psi_A}\longrightarrow \squet{\tilde{\Psi}_A}=   \int \mathcal{D}u ~
 c_u  e^{\theta(u, \varphi_{u\rightarrow\tilde{u}}*u)}  {\varphi_{u\rightarrow\tilde{u}}}^* \squet{ {\varphi_{u\rightarrow\tilde{u}}}^* u}
\end{equation}
This is a quantum diffeomorphism which transforms the points in $\mathcal{O}_u$ in each $\squet{u}$ term so that they coincidence with the points they are identified with in $\mathcal{O}_{\tilde{u}}$.   We can associate coordinates with $\mathcal{O}_{\tilde{u}}$ by means of the map
\begin{equation}
x_{\tilde{u}} (p)
\end{equation}
Since the $\mathcal{O}_u$ sets have been transformed to coincide with $\mathcal{O}_{\tilde{u}}$, this coordinate map associates these coordinates for all terms in the extended state (for the part of the manifold we are interested in).

Had we chosen $\tilde{u}'$ instead of $\tilde{u}$ we would get a different extended state.  We can map between these cases by applying the same diffeomorphism, $\varphi_{\tilde{u}\rightarrow\tilde{u}'}$ to each term of the expression on the left in \eqref{PsiAoverlap}.  We can write this as
\begin{equation}
\squet{\tilde{\Psi}'_A} = \varphi_{\tilde{u}\rightarrow\tilde{u}'} \squet{\tilde{\Psi}_A}
\end{equation}
Note this can be regarded as a classical diffeomorphism since every $\squet{u}$ term is subject to the same diffeomorphism.

Once we have a $\squet{\tilde{\Psi}_A}$ in this form (so the $\mathcal{O}_u$'s coincide) then we can apply a (restricted) quantum diffeomorphism to it,
\begin{equation}
\squet{\tilde{\Psi}_A} \longrightarrow \squet{\tilde{\Psi}'_A} = \squop{\varphi} \squet{\tilde{\Psi}_A}
\end{equation}
obtaining a new extended state now having the points for the sets $\mathcal{O}'_u = \varphi_u(\mathcal{O}_u)$ coinciding for different $u$ (here $\{\mathcal{O}_u\}$ are the maps associated with the quantum diffeomorphism).  This will change the identification between points in the manifold associated with different $u$ thus constituting a quantum coordinate transformation.  The new coordinates are given by
\begin{equation}
x_{\tilde{u}} (\varphi_{\tilde{u}}(p))
\end{equation}
Thus, if we take an active point of view then we see that quantum diffeomorphisms on the extended state are associated with quantum coordinate transformations.

One point that is worth making is the following.  A point, $x$ , in a quantum coordinate system is associated with a set of points, one in each manifold, $\mathcal{M}_u$.  If we perform a classical diffeomorphism then $x$ is transformed to $x'$.  Under this classical transformation we do not break the identification map and so we might say that $x$ keeps its identity (and is merely relabeled by $x'$).   However, if we perform a quantum diffeomorphism, then $x$ may lose its identity in the sense that it is no longer associated with any $x'$ in the new quantum coordinate system.  This is because quantum diffeomorphisms break the identification between points in the different manifolds.

It is instructive to consider an example in which a quantum coordinate transformation can be used to transform from a coordinate system in which one quantity is definite to one in which another is.  We can nominate a set of scalars, $\mathbf{S}=(S_1, S_2, \dots, S_K)$ (in \cite{hardy2016operational} such sets were used to set up an \lq\lq operational space").  Each scalar, $S_k(p)$, in this can be built out of the tensors in $\Phi$ and the action of the covariant derivative so that all indices are summed over.  We can set up a quantum coordinate system in which points are identified between different $\mathcal{M}_u$ that have the same $\mathbf{S}$.  We could consider this as going into the quantum frame of reference \lq\lq co-moving" with $\mathbf{S}$ (note, however, this frame of reference will not be unique if, for some $u$, more than one point, $p\in\mathcal{M}_u$, is mapped to the same $\mathbf{S}$).   We might consider a different set of scalars, $\mathbf{S}'=(S'_1, S'_2, \dots, S'_{K'})$.  We can, instead, consider a quantum coordinate system in which these points are identified.  When we are in the first quantum coordinate system we expect $\mathbf{S}'$ to be indefinite (so points identified with the same $x$ have different values for $\mathbf{S}'$).  Similarly, when we are in the second quantum coordinate system, we expect $\mathbf{S}$ to be indefinite.  An appropriate quantum coordinate transformation will take us between these two cases.





\section{Implementing the QEP}

The Quantum Equivalence Principle (QEP) says that, in the vicinity of any point, we can always find a quantum coordinate system such that we have definite causal structure.  By  \lq\lq point" we mean any point that can be defined with respect to some quantum coordinate system.  Such a point is given by specifying some particular point, $p_1\in \mathcal{M}_{\tilde{u}}$, for some particular $u=\tilde{u}$ along with a set of identification maps $\{\varphi_{\tilde{u}\rightarrow u}\}$ which identifies $p_1$ with a point $\varphi_{\tilde{u}\rightarrow u}(p_1)$ for every other $\mathcal{M}_u$.   Consider the extended $\mathbf{g}$-state
\begin{equation}
\squet{\Psi_\mathbf{g}} = \int \mathcal{D}u ~ c_u \mathbf{g}_u(\mathcal{M}_u) \squet{u}
\end{equation}
where $\mathbf{g}$ is the metric tensor (considered as an abstract tensor).  To implement the QEP we simply need to find a quantum diffeomorphism such that the metrics $\mathbf{g}_u$ are equal in the vicinity of the point in question with regard to the quantum coordinate system.   We will describe one way to do this.  We can act on $\squet{\Psi_\mathbf{g}}$ with a quantum diffeomorphism to bring the points identified with $p_1$  (for the other $u$) into coincidence (as discussed in Sec.\ \ref{sec:quantumcoordinatetransformations})
\begin{equation}
\squet{\tilde{\Psi}_\mathbf{g}} = \squop{\varphi} \squet{\Psi_\mathbf{g}}
\end{equation}
where $\squop{\varphi}$ is associated with the set $\{ \varphi_{u\rightarrow v}\}$ of diffeomorphisms that bring the points corresponding to $p_1$ into alignment along with a set of points in the vicinity (in associated $\mathcal{O}_u$ sets).   At this stage the metric living on the different $\mathcal{M}_u$ can be very different in the vicinity of $p_1$.  Now we perform a second quantum diffeomorphism, $\squop{\varphi}'$, which leaves $p_1$ unchanged.  The new extended $\mathbf{g}$-state is
\begin{equation}
\squet{\tilde{\Psi}'_\mathbf{g}} = \squop{\varphi}' \squet{\tilde{\Psi}_\mathbf{g}}
\end{equation}
where this quantum diffeomorphism is associated with a set, $\{\varphi_u\}$, of diffeomorphisms chosen such that now the metrics $\mathbf{g}_u$ are equal to first order in the vicinity of $p_1$. Thus, if we write
\begin{equation}
\squet{\tilde{\Psi}'_\mathbf{g}} = \int \mathcal{D}u ~ \tilde{c}'_u \tilde{\mathbf{g}}'_u(\tilde{\mathcal{M}}'_u) \squet{u}
\end{equation}
then we require
\begin{equation}\label{alignedgs}
\tilde{\mathbf{g}}'_u(x_1+\delta x) = \tilde{\mathbf{g}}'_v(x_1+\delta x) + O(\delta x^2)   ~~~~ \forall ~u~\text{and}~v ~s.t.\ ~ c_u\not=0~\text{and}~c_v\not=0
\end{equation}
We can think of the second quantum diffeomorphism ($\squop{\varphi}'$) as rotating the light cones associated with these different metrics (for the different $u$) so that they align and also stretching the manifolds so that the conformal factors associated with the metrics agree.  If there is a time direction (past to future) at each $p$ then we can rotate the light cones so that they agree on future and past.  We can always satisfy \eqref{alignedgs} because we can always transform the metric in the vicinity of any point to the Minkowski metric.

The transformation does a little more than implied by the quantum equivalence principle. In addition to aligning causal structure (light cones) it also matches the conformal factors.  We could relax the latter imposition so that we have a statistical mixture over different conformal factors (corresponding to different clock rates at $x_1$ for the different manifolds.  However, we have enough freedom to actually align these conformal factors so we will assume that we do this in what follows (we might call the fact that we can do this the \emph{strong quantum equivalence principle}).


\section{General Relativity}\label{sec:generalrelativity}

\subsection{The problem of Relativistic Gravity}

After Special Relativity had been discovered by Einstein in 1905 along with the space-time picture of Minkowski from 1907 it was understood that physical theories should be formulated as Special Relativistic Field Theories.  Maxwell's theory  of electromagnetism, already published some years earlier in 1861, was one such theory.  Fluid dynamics and other theories would subsequently be given formulations as Special Relativistic Field Theories.  However, in 1907, Newton's theory of Gravity resisted efforts to formulate it in these terms.   Thus, the stage was set for consideration of the following problem.
\begin{quote}
\textbf{The problem of Relativistic Gravity} is to find a physical theory that reduces in appropriate limits to the theory of Newtonian Gravity on the one hand and to Special Relativistic Field Theories (SRFT) on the other.
\[  \text{Newtonian Gravity} \longleftarrow \text{Relativistic Gravity} \longrightarrow \text{SRFT}   \]
\end{quote}

\subsection{How Einstein solved the problem of Relativistic Gravity}

Einstein solved the problem of Relativistic Gravity in the form of General Relativity.  His starting point was a realisation he had in 1907 - namely the equivalence principle.  He called this the \lq\lq happiest thought in my life" \cite{pais1982subtle}.    The equivalence principle acts as a bridge between the old physics and the new physics.  It is the essential clue in working out how to reverse the arrows so we have
\[  \text{Newtonian Gravity} \longrightarrow \text{Relativistic Gravity} \longleftarrow \text{SRFT}   \]
While the equivalence principle formed the right starting point, there was still much work to be done to go from this realisation to the full theory of General Relativity (and it took him until 1915 to complete this task).  It is worth outlining in some detail how Einstein did this.

First, let us be clear about what General Relativity is.  The theory is captured by \emph{three elements}\cite{hardy2018construction}
\begin{description}
\item[A prescription] for converting the field equations in a Special Relativistic Field Theory into general relativistic field equations.  This prescription (sometimes called \emph{minimal substitution})works as follows.
    \begin{itemize}
    \item The coordinates, $x^{\bar{\mu}}$, of the global inertial reference frame are replaced by general coordinates, $x^\mu$. This is done by replacing all indices  $\bar{\mu}$, $\bar{\nu}$, \dots by $\mu$, $\nu$, \dots
    \item The Minkowski metric, $\eta_{\mu\nu}$ is replaced by the general metric, $g_{\mu\nu}$.
    \item All partial derivatives, $\partial_\mu$ are replaced by covariant derivatives, $\nabla_\mu$.
    \end{itemize}
    The field equations obtained in this way are called the matter field equations.
\item[An addendum.]  The general metric $g_{\mu\nu}$ introduces an additional ten real parameters into the theory (for 4-dimensional spacetime).  Thus, to have a complete set of field equations we need (it would seem) an additional ten field equations.  Einstein provided just such a set in the form of the Einstein field equations
    \[    G^{\mu\nu} = 8\pi T^{\mu\nu}   \]
    These equations satisfy a form of energy momentum transformation (namely $\nabla_\mu T^{\mu\nu}=0$) because of the mathematical identity $\nabla_\mu G^{\mu\nu} =0$.  However, this means that Einstein's field equations actually only furnishes us with six independent field equations (since the mathematical identity shows that four of them are related).  It would seem that we do not have a complete set of field equations after all.  In fact we do not need extra equations because of the next element.
\item[An Interpretation.]  The beables (physically real quantities) are those quantities that are invariant under diffeomorphisms.  To elaborate, we can regard a solution, $u$, to the field equations (both the matter and Einstein field equations) as a set
    \[ u= \left\{ (\Phi, p): \forall p\in \mathcal{M} \right\}    \]
    that specifies the values of all the tensor fields (denoted by $\Phi$) at each point, $p$, in some manifold, $\mathcal{M}$.
\end{description}
The need for the second element follows from the first. The need for the third element follows from the missing equations in the second and is, in any case, necessitated by the fact that all the equations are invariant under general coordinate transformations.  Thus, acting on a solution with a general coordinate transformation produces another valid solution. To see how this works, consider the following.  When we represent the points, $p\in\mathcal{M}$  with coordinates, $x^\mu$, a diffeomorphism corresponds to a coordinate transformation represented by four equations: $x^\mu \rightarrow x^{\mu'}(\{ x^\mu\})$.   These four equations correspond to the missing equations of the second element.

\begin{figure}
  \centering
\begin{tikzpicture}[xscale=0.7]
\node[bigboxP] (principles){Principles};
\node[smallbox] (equivalence) at ([yshift=-3.5em]principles.north){Equivalence principle};
\node[smallbox] (general) at ([yshift=-0.25em]principles.center){General covariance};
\node[smallbox] (conservation) at ([yshift=2.5em]principles.south){Energy-momentum conservation};
\node[bigboxE]  (elements)  at  ([yshift=-4cm]principles.south) {Elements of GR};
\node[smallbox] (prescription) at ([yshift=-3em]elements.north){Prescription};
\node[smallbox] (addendum) at ([yshift=-0.75em]elements.center){Addendum};
\node[smallbox] (interpretation) at ([yshift=1.5em]elements.south){Interpretation};
\path[draw,->] (prescription) to (addendum); \path[draw,->] (addendum) to (interpretation);
\node[bigboxM]  (maths)  at  ([xshift=5cm]principles.east) {Mathematics};
\node[smallbox] (coords) at ([yshift=-3.5em]maths.north){General coordinates};
\node[smallbox] (manifolds) at ([yshift=-2em]coords.south){Manifolds};
\node[smallbox] (tensors) at ([yshift=-2em]manifolds.south){Tensor fields};
\node[smallbox] (derivative) at ([yshift=-2em]tensors.south){Covariant derivative};
\path[draw,->] (coords) to (manifolds); \path[draw,->] (manifolds) to (tensors); \path[draw, ->] (tensors) to (derivative);
\node[bigboxT]  (theories)  at  ([yshift=-4cm]maths.south) {Old Theories};
\node[smallbox] (SRFT) at ([yshift=-4em]theories.north){Special Relativistic Field Theories};
\node[smallbox] (newton) at ([yshift=-2em]SRFT.south){Newtonian Gravity};
\FlowchartlineEWsmooth[8pt, 8pt]{equivalence}{coords}{2.5} %
\FlowchartlineWW[0pt, -3pt]{equivalence}{prescription}{2.9}
\FlowchartlineWW[-6pt, 3pt]{general}{prescription}{1.7}
\FlowchartlineWW[0pt,-3pt]{general}{addendum}{2.1}
\FlowchartlineWW[6pt,0pt]{general}{interpretation}{2.5}
\FlowchartlineEW[0pt, 0pt]{general}{tensors}{2.3}   %
\FlowchartlineEW[-5pt,8pt]{general}{derivative}{1.2}  %
\FlowchartlineWW[0,3pt]{conservation}{addendum}{1.3}
\FlowchartlineEE[0pt,0pt]{SRFT}{tensors}{1.5}
\FlowchartlineWE[5pt,3pt]{newton}{equivalence}{1.25}
\FlowchartlineWE[0pt,3pt]{newton}{addendum}{2}
\FlowchartlineWEsmooth[0pt, -1pt]{SRFT}{prescription}{2.5}
\FlowchartlineWE[10pt, -10pt]{SRFT}{conservation}{1.6}
\FlowchartlineWE[1pt, 2pt]{derivative}{prescription}{2}
\FlowchartlineWE[-1pt, 5pt]{coords}{general}{2.7}
\FlowchartlineWE[-7pt, 5.5pt]{coords}{prescription}{2.35}
\FlowchartlineEWred[-3pt,-5pt]{addendum}{newton}{1.5}
\FlowchartlineEWsmoothred[-5.5pt,-8pt]{prescription}{SRFT}{what}
\end{tikzpicture}
\caption{The elements of General Relativity are motivated by principles, mathematical structures, and the old physical theories. The black arrows indicate influences.  The red arrows indicate how the old physical theories can be obtained as a limiting case.}\label{fig:GR}
\end{figure}
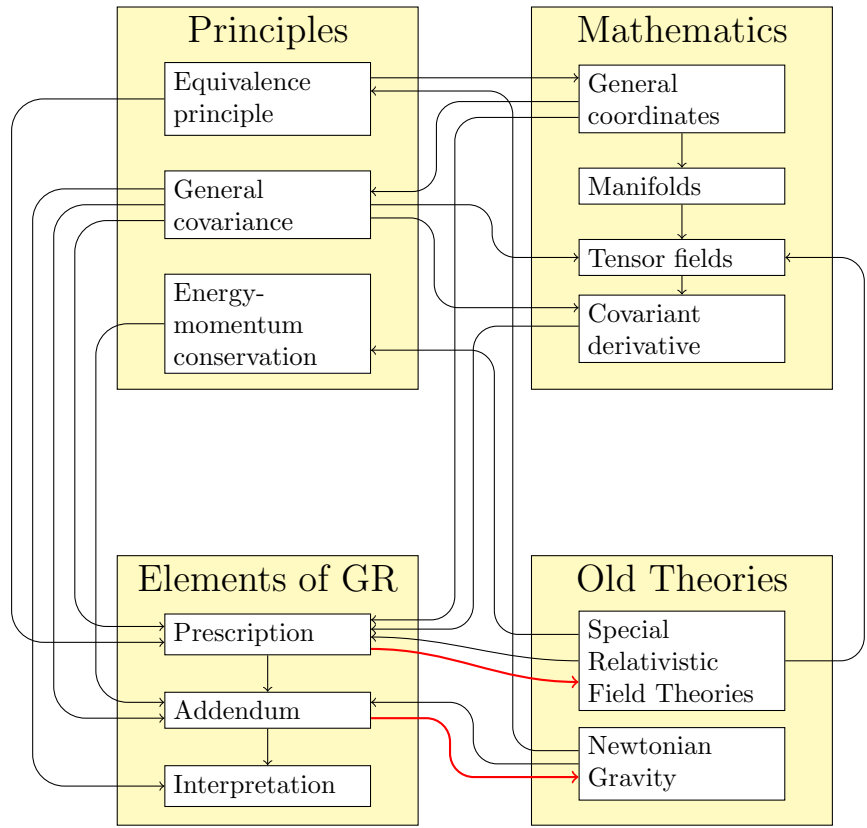

These three elements are motivated in a bigger environment employing principles, mathematical structures, and using the old theories of Newtonian Gravity and Special Relativistic Field Theories.  This is captured in Fig.\ \ref{fig:GR}.  The black arrows represent lines of influence.  For example, the equivalence principle is motivated by physics already in Newton's theory of gravity (of course, the principle actually goes back to Galileo's observation that different masses fall at the same rate).  The equivalence principle motivates the move to general coordinates which is the starting point in setting up the mathematical structure used in General Relativity.   The equivalence principle finds a particular role in the prescription (by replacing derivatives with covariant derivatives). The principle of general covariance states that the laws of physics should be written in a way that they take the same form in any coordinate system.  This motivates adopting the use of tensor fields in expressing physical laws.  Einstein was motivated by the Poison equation formulation of Newtonian Gravity to look for an equation that is second order in derivatives of the metric (since the metric is playing the role of Newton's gravitational potential).
This figure is not complete.  Additional principles could be included.  Mach's principle influenced Einstein's thinking but it is not clear that it plays a direct role in dictating the mathematical form of the elements of General Relativity.  The equivalence principle is closely related to the principle of local flatness (that we can always find a reference frame with respect to which the metric is Minkowski in the vicinity of any given point). These two principles become equivalent under the assumption of metric compatibility (that $\nabla_\mu g^{\mu\nu} =0$) \cite{hardy2016operational}.  The fact that we adopt a torsion free covariant derivative might warrant some sort of motivating principle (it follows from requiring that covariant derivatives, like regular partial derivatives, commute when acting on scalar fields). There is much additional mathematical structure not explicitly mentioned.  There are particular tensor fields that play an important role such as the metric tensor ($g_{\mu\nu}$), the energy-momentum tensor ($T^{\mu\nu}$), and the Einstein tensor ($G^{\mu\nu}$).  There are the notions of coordinate transformations and, relatedly, diffeomorphisms.  And there are the Bianchi identities.  All these objects might be represented in an enriched diagram.

One might argue for additional lines of influence.  For example, the mathematical notion of tensor fields is important for every element of General Relativity.  Also, we can argue that the principle of energy-momentum conservation is motivated by Newtonian Gravity as well as Special Relativistic Field Theories. However, it is the version from SRFTs (that $\partial_\mu T^{\mu\nu}=0$) that finds particular application.  In any case, the diagram is already busy enough so less pertinent lines of influence have  been omitted.

The red arrows in Fig.\ \ref{fig:GR} indicate that General Relativity has appropriate limits to Newtonian Gravity and to Special Relativistic Field Theories  - thus solving the problem of Relativistic Gravity. It is interesting that General Relativity modifies both the old theories in order to do this - there is, perhaps, a lesson for Quantum Gravity in this.

\section{Quantum Gravity}\label{sec:quantumgravity}

\subsection{The problem of Quantum Gravity}

Today we face a problem that is analogous to the problem of Relativistic Gravity.  We have two physical theories that are each successful in their own realms (General Relativity and Quantum Theory) but they do not fit together.   The stage is set for consideration of the following.
\begin{quote}
\textbf{The problem of Quantum Gravity} is to find a theory that reduces in appropriate limits to General Relativity on the one hand and to Special Relativistic Quantum Field Theory (SRQFT) on the other.
\[ \text{General Relativity} \longleftarrow \text{Quantum Gravity} \longrightarrow \text{SRQFT}   \]
\end{quote}
Special Relativistic Quantum Field Theory seems like the appropriate version of Quantum Theory for the limiting case here.  Of course, there remain technical, structural, and conceptual questions about formulating Quantum Field Theories and we might even hope that we gain some insight into these by solving the problem of Quantum Gravity (for example, Ding Jia has suggested that indefinite causal structure may offer a route to regularising Quantum Field Theory \cite{jia2018reduction}).

\subsection{A proposed path to a theory of Quantum Gravity}

\begin{figure}
\centering
\begin{tikzpicture}[xscale=0.7]
\node[bigboxQP] (principles){Principles};
\node[smallbox] (equivalence) at ([yshift=-4em]principles.north){Quantum equivalence principle};
\node[smallbox] (general) at ([yshift=-3em]equivalence.south){General quantum covariance};
\node[smallbox] (conservation) at ([yshift=-3em]general.south){Energy-momentum conservation};
\node[bigboxE]  (elements)  at  ([yshift=-4cm]principles.south) {Elements of QG};
\node[smallbox] (prescription) at ([yshift=-3em]elements.north){Prescription};
\node[smallbox] (addendum) at ([yshift=-0.75em]elements.center){Addendum};
\node[smallbox] (interpretation) at ([yshift=1.5em]elements.south){Interpretation};
\path[draw,->] (prescription) to (addendum); \path[draw,->] (addendum) to (interpretation);
\node[bigboxQM]  (maths)  at  ([xshift=5cm]principles.east) {Mathematics};
\node[smallbox] (coords) at ([yshift=-3.5em]maths.north){General quantum coordinates};
\node[smallbox] (manifolds) at ([yshift=-2em]coords.south){Quantum manifolds};
\node[smallbox] (tensors) at ([yshift=-2em]manifolds.south){Quantum tensor fields};
\node[smallbox] (derivative) at ([yshift=-3em]tensors.south){Quantum covariant derivative};
\path[draw,->] (coords) to (manifolds); \path[draw,->] (manifolds) to (tensors); \path[draw, ->] (tensors) to (derivative);
\node[bigboxT]  (theories)  at  ([yshift=-4cm]maths.south) {Old Theories};
\node[smallbox] (SRFT) at ([yshift=-4em]theories.north){Special Relativistic Quantum Field Theories};
\node[smallbox] (newton) at ([yshift=-2em]SRFT.south){General Relativity};
\FlowchartlineEWsmooth[8pt, 8pt]{equivalence}{coords}{2.5} %
\FlowchartlineWW[0pt, -3pt]{equivalence}{prescription}{2.9}
\FlowchartlineWW[-6pt, 3pt]{general}{prescription}{1.7}
\FlowchartlineWW[0pt,-3pt]{general}{addendum}{2.1}
\FlowchartlineWW[6pt,0pt]{general}{interpretation}{2.5}
\FlowchartlineEW[0pt, 0pt]{general}{tensors}{2.3}   %
\FlowchartlineEW[-5pt,8pt]{general}{derivative}{1.2}  %
\FlowchartlineWW[0,3pt]{conservation}{addendum}{1.3}
\FlowchartlineEE[0pt,0pt]{SRFT}{tensors}{1.5}
\FlowchartlineWE[5pt,-10pt]{newton}{equivalence}{1.25}
\FlowchartlineWE[0pt,3pt]{newton}{addendum}{2}
\FlowchartlineWEsmooth[0pt, -1pt]{SRFT}{prescription}{2.5}
\FlowchartlineWE[10pt, -10pt]{SRFT}{conservation}{1.6}
\FlowchartlineWE[-8pt, 2pt]{derivative}{prescription}{2}
\FlowchartlineWE[1pt, 5pt]{coords}{general}{2.7}
\FlowchartlineWE[-5pt, 5.5pt]{coords}{prescription}{2.35}
\FlowchartlineEWred[-3pt,-5pt]{addendum}{newton}{1.5}
\FlowchartlineEWsmoothred[-5.5pt,-8pt]{prescription}{SRFT}{what}
\end{tikzpicture}
  \caption{We can consider attempting to construct a theory of Quantum Gravity by following a schema that is analogous to that used to obtain General Relativity.   There is much work to be done in elucidating what the different parts of this diagram might mean in this case. The present paper makes concrete proposals for some of the elements.}\label{fig:QG}
\end{figure}
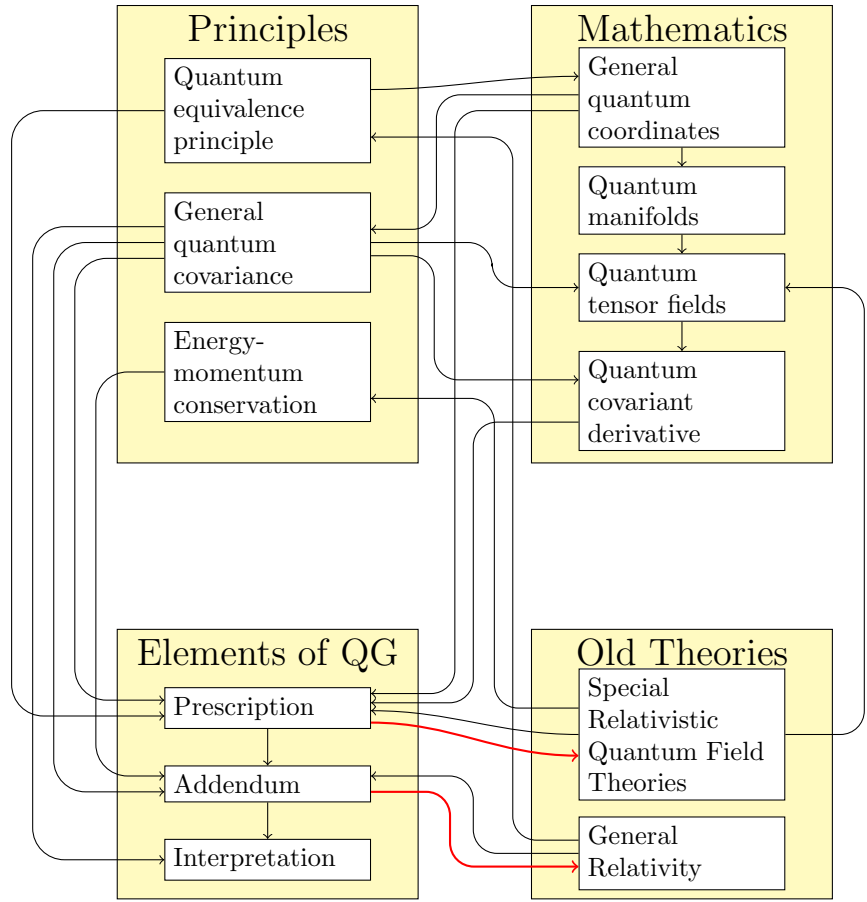
Can we take a path that is analogous to Einstein's path to solving the problem of Quantum Gravity?  One way to think about this is to appropriate Fig.\ \ref{fig:GR}.    This idea is ilustrated in Fig.\ \ref{fig:QG} which has been converted from Fig.\ \ref{fig:GR} by changing the \lq\lq old theories" to Special Relativistic Quantum Field Theories and to General Relativity and inserting the word \lq\lq quantum" where appropriate.

Proposals for some elements of Fig.\ \ref{fig:QG} have been provided already in this paper.   We have a quantum equivalence principle.  We have a notion of quantum coordinates.   We have a notion of quantum diffeomorphisms which clarifies what the \emph{interpretation} means.  The principle of general quantum covariance can be extrapolated from the classical principle:
\begin{quote}
\emph{The principle of general quantum covariance} says that the laws of physics can (and should) be written in such a way that they take the same form in any quantum coordinate system
\end{quote}
A similar idea was suggested by Giacomini, Castro-Ruiz, and Brukner \cite{giacomini2019quantum}.   They set up a notion of quantum reference frames at a given time (so they are in space but not in space-time) and posited that physical laws should be covariant with respect to transformations between such frames.  In fact they go a bit further and prove that the Schroedinger equation is covariant.
 We have discussed the interpretation element of Quantum Gravity in Sec.\ \ref{sec:beables}.  The notion of an extended $A$-state does not seem to quite do the job of tensors in General Relativity so some work is necessary there (we will make some proposals for this in Sec.\ \ref{sec:quantumtensorfields}).   We do not have a suitable notion of a quantum covariant derivative but there is reason to be hopeful that this can be built given the notions we do have.

\subsection{How can we use the QEP?}

Now that we have the quantum equivalence principle, how should we use it to construct a theory of Quantum Gravity? We can get some suggestions from Fig.\ \ref{fig:QG}.   There are two lines of influence coming out of the quantum equivalence principle in Fig.\ \ref{fig:QG}: (i) one goes to general quantum coordinates; (ii) the other goes to the prescription.   We have already shown how we can set up a notion of general quantum coordinates motivated by the QEP.  What about the prescription? In General Relativity the equivalence principle is used to convert special relativistic field equations to general relativistic ones in such a way that we maintain the property that there exists an inertial coordinate system in the local vicinity of any given point.   To do this requires a notion of covariant derivative.  Thus, in the quantum context, we wish to find a way to convert special relativistic quantum field equations for the matter degrees of freedom into quantum gravity field equations in such a way that we maintain the property that there exists a causally definite quantum coordinate system.  Aligning the metric for the different $\squet{u}$ terms by applying a quantum diffeomorphism will not, in general, also align the matter fields.  However, this will allow us to implement \emph{causality} conditions locally at any given point.  This seems to be the most promising avenue for research.

It is worth noting that causality is not listed among the principles of Fig.\ \ref{fig:GR}. This is because Einstein did not make explicit use of causality in setting up the theory of General Relativity.  Nevertheless, General Relativity is causal.  Local disturbances cannot propagate outside the light cone determined by the metric.   This is particularly interesting in the context of General Relativity because we can use disturbances in the metric itself (the very object that determines what we mean by causality) to send signals.   It is, then, striking that causality comes out of General Relativity given that it was not put in.  It would be interesting to pursue a different route to General Relativity wherein causality was explicitly used as a principle in the derivation of the theory.    If we could do this in the classical context, then it might shed light on how to use causality as a principle in obtaining Quantum Gravity exploiting the quantum equivalence principle along the way.

\subsection{Quantum manifolds}\label{sec:quantummanifolds}

I will now outline some \emph{tentative} ideas in which we make sense of a notion of quantum manifolds with quantum tensor fields on them.   First note that, a (classical) tensor field is specified by giving its components at each point, $p$, in a manifold.  The point, $p$, has no intrinsic physical properties of  its own (physical properties only emerge  through relational properties between the various fields defined at $p$).  Let us recall a few definitions we have used so far.  An extended $\mathbf{A}$-state is written
\begin{equation}
\squet{\Psi_\mathbf{A}} = \int \mathcal{D}u ~ c_u\mathbf{A}_u \squet{u}
\end{equation}
The set, $\mathbf{A}_u$, is defined as
\begin{equation}
\mathbf{A}_u = \left\{ (\mathbf{A}(p), p): \forall p\in\mathcal{M}_u \right\}
\end{equation}
Here $u$ is
\begin{equation}
u= \left\{ (\Phi(p), p): \forall p\in\mathcal{M}_u \right\}
\end{equation}
Thus, $u$, contains full information about the given configuration of classical fields.  The idea I want to propose here is to disregard this aspect of $u$ and regard it as indexical - that is regard it as being of a similar status to a point $p$ in a manifold, $\mathcal{M}$.  This is a big step as $u$ is already a set of fields defined on a manifold. However, we can think of $u$ as being represented a point in its own space, $\mathcal{S}$.    We can take the following steps to implement this. First we disallow any operations that act on $\squet{u}$ to bring out tensor fields from $\Phi$.  Second, we can imagine \lq\lq scrambling" or \lq\lq forgetting" the physical information in $u$ so we are just left with the topological information. We could do this by applying some arbitrary diffeomorphism.

We do, however, need to keep three pieces of information.    First we need to keep $\mathcal{M}_u$.  Second, we need to kep the information as to whether or not two elements, $u$ and $v$ in $\mathcal{S}$ can be transformed into one another by a diffeomorphism (which we denote $u \simeq v$).  This enables us to apply quantum diffeomorphisms.  Third, we need to keep the partial order $u\subseteq v$ so we know how to evaluate $[u\shortlongverttextmode v]$.  We can use $\mathcal{S}$ and $\mathcal{M}_u$ to build a space which we will call a \emph{quantum manifold}.

A quantum manifold is an element, $\mathcal{Q}$, in a \emph{quantum fibre bundle} which will be defined  below.  First we define the simpler notion of a \emph{basic quantum fibre bundle} (this does not include the $u\simeq v$ and $u\subseteq v$ structure) which is modeled on the definitions of a fibre bundle and a manifold.  We have a base space, $\mathcal{S}$, and at each point, $u\in \mathcal{S}$ is attached a manifold, $\mathcal{M}_u$.  We impose what we call the local-local triviality condition intended to ensure that we can smoothly sew the quantum manifold together.
\begin{quote}
\textbf{A basic quantum fibre bundle} is defined by
\[   (\mathcal{Q}, \mathcal{S}, \pi, M)    \]
such that:
\begin{itemize}
\item for each $q\in \mathcal{Q}$, $\pi(q)\in \mathcal{S}$;
\item for each $u\in \mathcal{S}$, $M(u)=\mathcal{M}_u$ where $\mathcal{M}_u$ is a manifold (recall that this can have corners according to our usage of the term).
\item there exists a bijection, $h$, between points $q\in \mathcal{Q}$ and $(u, p)$ where $u=\pi(q)$ and $p\in \mathcal{M}_u$.
\[ h(q) = (u,p)  \]
\end{itemize}
Further we require that the following \emph{local-local triviality conditions} hold:
\begin{itemize}
\item  $\mathcal{Q}$ is covered by open sets, $\mathcal{R}_i$,  so that
\[ \bigcup_i \mathcal{R}_i = \mathcal{Q}  \]
\item there exists a bijection, $g_i$, such that
\[ g_i(\mathcal{R}_i) = \pi(\mathcal{R}_i)\times V   \]
and where $V\subseteq \mathbb{R}^N$;
\item the sets,  $\mathcal{O}_{ui} \subseteq \mathcal{M}_u$, defined by
\[   p\in \mathcal{O}_{ui} ~~ \text{iff} ~~  h^{-1}(u,p) \in \mathcal{R}_i   \]
are open and form a cover for $\mathcal{M}_u$ so that
\[ \bigcup_i \mathcal{O}_{ui} = \mathcal{M}_u   \]
\item the maps $\omega_{ui}(p)$ from $p\in \mathcal{O}_{ui}$ to $x\in V$ defined by
\[ \omega_{ui}(p) =    \text{proj}_2  g( h^{-1}(u, p)   )  \]
satisfies the usual axioms for such a cover for a manifold.  These are:  (i) for points $p\in \mathcal{O}_{uj}\cap\mathcal{O}_{ui}$ we require $\omega_{uj}(\omega_{ui}^{-1}(x))$ is smooth; and (ii) $\omega(\mathcal{O}_{uj}\cap\mathcal{O}_{ui})$ is open.  Note that $\text{proj}_2$ projects onto the second factor in the cartesian product $\pi(\mathcal{R}_i)\times V$ returning an $x\in V$.
\end{itemize}
\end{quote}
The manifolds, $\mathcal{M}_u$, may be topologically distinct from one $u$ to another (whereas the fibres in a standard fibre bundle are all topologically equivalent).  However, the local spaces, $\mathcal{O}_{ui}$, are topologically equivalent to $V$ (for all $u
$).  Hence we choose to build a something like fibre bundle with fibres $V$ while ensuring that the $\mathcal{O}_{ui}$ sets can be sewed together to form a manifold at each $u$.  More attention is required to what types of maps $h$ and $g_i$ are.  Above they are described as bijections. However, we may want to impose continuity and, even, smoothness on them also.  Certainly they  need to be smooth enough that the $\mathcal{O}_{ui}$ sets satisfy the smoothness condition given.  However, given that the manifolds, $\mathcal{M}_u$, at different $u$ may be topologically different, these maps may also have to have discontinuities.   Assuming we can count the distinct topologies, we may correspondingly be able to impose the constraint that $h$ and $g$ have a countable set of discontinuities and are otherwise smooth and still have a useful definition for quantum manifolds.  One of the two local's in \lq\lq local-local triviality condition" comes from the local sewing together of the $\mathcal{O}_{ui}$'s and the other from the usual local triviality condition for fibre bundles.   We leave for future work to prove whether or not $\mathcal{Q}$ is actually a manifold in the usual technical sense. However, it deserves the name \lq\lq quantum manifold" because it is the quantum analogue of a manifold as used in (classical) General Relativity.

The functions  $g_i(\cdot)$ and $h(\cdot, \cdot)$ can be used to implement a quantum coordinate system that covers  those $u\in \pi(\mathcal{R}_i)$.  The quantum coordinate associated with $q\in\mathcal{R}_i$ is
\begin{equation}\label{qcoordmap}
x_i(q) :=  \text{proj}_2 g_i(q)
\end{equation}
(note that the $i$ subscript indicates that this coordinate charts $\mathcal{R}_i$).  We can use $g_i$ and $h$ to construct an identification map from the points in $\mathcal{O}_{ui}$ to the points in $\mathcal{O}_{vi}$ for the quantum coordinate system as follows:
\begin{equation}
\varphi_{u\rightarrow v} =    \text{proj}_2 h(    g_i^{-1}(v,  x_i( h^{-1}(u, p) )  )                  )
\end{equation}
This maps a point $p\in \mathcal{O}_{ui}$ to a point $q\in \mathcal{Q}$ using $h^{-1}$.  Then it uses $x_i$ to map this to some coordinate $x$.  Then, using $g_i^{-1}$, it maps this coordinate to a point $q\in\mathcal{Q}$ for which $\pi(q)=v$.  Using $h$ we map to a point $(v,p)$ where $p\in \mathcal{O}_{vi}$.  Then finally, we project on to the second entry using $\text{proj}_2$ and obtain $p\in \mathcal{O}_{vi}$.  Since we have a set of quantum coordinate systems (or quantum charts) that cover the whole of the quantum manifold (as we vary over $i$) we have a quantum atlas in the sense discussed at the end of Sec.\ \ref{sec:quantumcoordinates}.

We call the above a \emph{basic} quantum fibre bundle since, as stated earlier, we actually need to add extra structure corresponding to the fact that elements $u$ and $v$ of $\mathcal{S}$ have two relationships between them in the physics we have considered that we would like to keep at an abstract level (that is, without explicit reference to the physical fields, $\Phi$, defined on the manifolds $\mathcal{M}_u$ and $\mathcal{M}_v$).   First, two such elements may be diffeomorphism equivalent (either can be transformed into the other by a diffeomorphism).   We denote this by $u \simeq v$. This is a transitive relationship (if $u\simeq v$ and $v\simeq w$ then $u\simeq w$).   Second, one such element may contain the other which we denote by $u\subseteq v$.  This is a partial order.  This gives us the full definition we seek
\begin{quote}
\textbf{A quantum fibre bundle} is defined by
\[   (\mathcal{Q}, \mathcal{S}, \pi, M, \simeq, \subseteq)    \]
such that it is a basic quantum fibre bundle in the elements $ (\mathcal{Q}, \mathcal{S}, \pi, M)$ and further:
\begin{itemize}
\item $\simeq$ is a transitive relationship that can hold on pairs of elements $u, v \in \mathcal{S}$ such that
\[ u\simeq v ~~~ \Rightarrow  ~~~\mathcal{M}_v= \varphi(\mathcal{M}_u) ~~ \text{for some} ~~ \varphi  \]
\item $\subseteq$   is a partial order on elements in $\mathcal{S}$ such that
\[   u\subseteq v  \Rightarrow \mathcal{M}_u \subseteq \mathcal{M}_v      \]
\item the following completeness condition holds
\[ \text{if}~~ \exists ~u, v, w \in \mathcal{S} ~~ \text{s.t.} ~~ u\subseteq v \simeq w ~~ \text{then} ~~\exists ~  y\in \mathcal{S} ~~\text{s.t.} ~~  u\simeq  y \subseteq w   \]
\end{itemize}
\end{quote}
Now we have enriched these quantum manifolds with the additional abstract structure required to be able to do those manipulations on $\squet{u}$ we discussed earlier  -  evaluating $[u\shortlongverttextmode v]$ and applying quantum diffeomorphisms.  The conditions given on manifolds for when $u\simeq v$ or $u\subseteq v$ are necessary but not sufficient conditions (as indicated by the $\Rightarrow$) because, in the actual physical examples, the fields defined on these manifolds also have to bear a certain relationship for $\simeq$ or $\subseteq$ to hold. The completeness condition expresses the idea that if $u$ is contained in $v$ and $v$ is transformed  to $w$  by $\varphi$ then there should exist some element $y$ that is obtained by acting on $u$ with $\varphi$.

\subsection{Quantum Tensor Fields}\label{sec:quantumtensorfields}

We can specify a \emph{quantum tensor field} by providing a tensor
\[ \mathbf{A}(q)     \]
at each point $q\in\mathcal{Q}$ where this tensor has components defined with respect to the tangent space of $\mathcal{M}_{\pi(q)}$ (thus the indices, $\mu$, from 1 to $\text{dim}(\mathcal{M}_u)$).

We can lay down a quantum coordinate system, $x$, in some region $\mathcal{R}\subseteq \mathcal{Q}$ as explained at the end of Sec.\ \ref{sec:quantummanifolds}.  Then we can represent the quantum tensor field explicitly by giving components
\[    A_{\mu\nu\dots}^{\gamma\delta \dots} (x, u)   \]
where $u=\pi(q)$.   We can sum over indices in the usual way. For example,
\[   D_\nu=  B_{\mu\nu}^{\gamma}C_\gamma^\mu   \]
Here we perform the summation at each point $(x, u)$.

We can use a quantum tensor field to define an extended state
\begin{equation}
\squet{\Psi}  = \int \mathcal{D}\pi(q) ~  \left\{ (\mathbf{A}(q), p): \forall p\in \mathcal{M}_{\pi(q)}  \right\} \squet{\pi(q)}
\end{equation}
Thus, we can go from quantum tensor fields to the path integral and recover quantum predictions in the usual way.


\section{Questions, comments and conclusions}

In this paper we have found a way to implement the quantum equivalence principle I originally proposed in \cite{hardy2018construction}.  The main idea is that a quantum coordinate system sets up an identification between the different manifold points corresponding to different terms in a quantum superposition and then associated coordinates, $x$, with the points.   A quantum diffeomorphism can break this identification and so corresponds to a bigger class of symmetries than simple classical diffeomorphisms.  We showed how we can use a quantum diffeomorphism to implement the quantum equivalence principle.   We then discussed the conceptual structure of General Relativity (shown in Fig.\ \ref{fig:GR}) and conjectured that a similar structure (shown in Fig.\ \ref{fig:QG}) may be possible for  Quantum Gravity using the quantum equivalence principle.

There are a number of questions that emerge from this project.  First, it is not clear that this is the only (or indeed the best way) to implement the idea of quantum coordinate systems for implementing the quantum equivalence principle.  In particular, it is not clear that the ideas here can  be used to account for the (albeit discrete) situation considered by Guerin and Brukner \cite{guerin2018observer} and by Oreshkov \cite{oreshkov2018whereabouts}.

It would be interesting to explore the connections between the approach to quantum coordinates considered here and quantum reference frames as considered by Giacomini, Castro-Ruiz, and Brukner (GCB).   Here are a few features of the GCB approach that are not features of the approach here.    First, the GCB approach uses a $3+1$ split into space and time.  Second, the approach of GCB places the reference frame on a particular quantum system.  Third,  the GCB approach eliminates the state associated with the quantum system the quantum reference frame is associated with (though it seems that they could keep this part of the state without affecting the main points they make).

A major issue we have not resolved is fixing the phase function, $\theta(u, v)$, used in defining quantum diffeomorphisms (see Sec.\ \ref{sec:quantumdiffeomorphisms}).    One possibility is that we simply set it equal to zero.  However, there may be physical motivations for some other choice.  In particular, this phase function may be fixed by considering the relationship between the approach to quantum coordinates presented here and the GCB approach to quantum reference frames (in which a phase is acquired on performing a quantum reference frame transformation).

We have not developed a theory of measurement for extended $A$-states beyond saying that they can be used to calculate amplitudes via the path integral.   We may be able to do much more.  First, we can select on $A$-states by demanding that they have certain beable properties (these beable properties corresponding to the outcomes) as discussed in Sec.\ \ref{sec:beables}.  We could use operational space corresponding to coincidences in the values of some set of scalars (as defined in \cite{hardy2016operational}) for this purpose.   Second, we may be able to take ratios of functions formed on $A$-states to determine whether probability ratios for different such outcomes are well defined (independent of choices elsewhere) and, if they are, what these probabilities are equal to (this is the formalism locality approach outlined in \cite{hardy2010bformalism}).  This kind of approach may take us beyond the path integral way of calculating probabilities.

To push forward this project we need to say what object is analogous to the covariant derivative.  In General Relativity, the covariant derivative replaces the partial derivative in field equations (this happens in the prescription).  We need to find an analogous procedure for  quantum field equations.

\section*{Acknowledgements}

I am deeply grateful to Flaminia Giacomini, Esteban Castro-Ruiz, and Philippe Gu\'erin explaining to me their work as well as broader discussions.  I am grateful to \v{C}aslav Brukner for email correspondence.   I am also very grateful to Ding Jia for explaining his superposed spacetimes approach and for comments on an earlier version of this paper.   I am grateful to Joy Christian for discussions on fibre bundles and the role non-trivial geometries may play in Quantum Gravity.  I am grateful to Zivy Hardy for support and the people at Aroma Cafe where most of this work was done.

Research at Perimeter Institute is supported by the Government of Canada through Industry Canada and by the Province of Ontario through the Ministry
of Economic Development and Innovation.  This project was made possible in part through the support of a grant from the John Templeton Foundation. The opinions expressed in this publication are those of the author and do not necessarily reflect the views of the John Templeton Foundation.  I am grateful also to FQXi for support through the grant FQXi-RFP-1824 entitled \lq\lq Operationalism, Agency, and Quantum Gravity".

\bibliography{QGbibJuly2019}
\bibliographystyle{plain}

\end{document}